\documentclass[preprint]{ptephy_v1}

\def\b8{$^8{\rm B}$}
\def\be7{$^7{\rm Be}$}

\newcommand{\nue}       {{\nu}_{\rm e}}

\newcommand{\gsim}      {\mathrel{\rlap{\raisebox{0.3ex}{$>$}}\raisebox{-0.6ex}{$\sim$}}}
\newcommand{\lsim}      {\mathrel{\rlap{\raisebox{0.3ex}{$<$}}\raisebox{-0.6ex}{$\sim$}}}
\def\nue{\nu_{e}}
\def\num{\nu_{\mu}}

\def\nmnt{$\nu_{\mu}\leftrightarrow\nu_{\tau}$~}
\def\nenm{$\nu_{e}\leftrightarrow\nu_{\mu}$~}

\def\lsim{\lower.7ex\hbox{${\buildrel < \over \sim}$}}
\def\gsim{\lower.7ex\hbox{${\buildrel > \over \sim}$}}

\begin{document}

\title{Toward the confirmation of atmospheric neutrino oscillations}


\author{Yuichi Oyama \\
\small
  High Energy Accelerator Research Organization (KEK), \\
  Oho 1-1, Tsukuba, Ibaraki 305-0801, Japan\\
  and \\
  J-PARC Center,
  Shirakata 2-4, Tokai, Ibaraki 319-1195, Japan
  \email{yuichi.oyama@kek.jp}}
\normalsize


\begin{abstract}%
The atmospheric muon neutrino deficit, which was possible evidence of \nmnt oscillation,
was reported by the Kamiokande experiment from 1988.
Many experimental efforts were made to examine the Kamiokande results.
Experiments which contributed to the confirmation of \nmnt oscillation are reviewed.
Especially, long-baseline neutrino-oscillation experiments
are described in detail.
\end{abstract}

\subjectindex{C04, C32}

\maketitle

\section{Introduction}
The atmospheric muon neutrino deficit, which might be evidence
of neutrino oscillations, was initially reported by the Kamiokande
experiment. The collaboration published three atmospheric
neutrino papers \cite{KAMATM1,KAMATM2,KAMATM3} in 1988, 1992 and 1994.
Details about the three Kamiokande papers
were reported in Ref. \cite{KAJITA2023}.
The history of the birth of the Kamiokande experiment
with the strong leadership of Prof. Masatoshi Koshiba was given in Ref. \cite{SUZUKI}.

\begin{figure}[b!]
\begin{center}
\includegraphics[width=12cm]{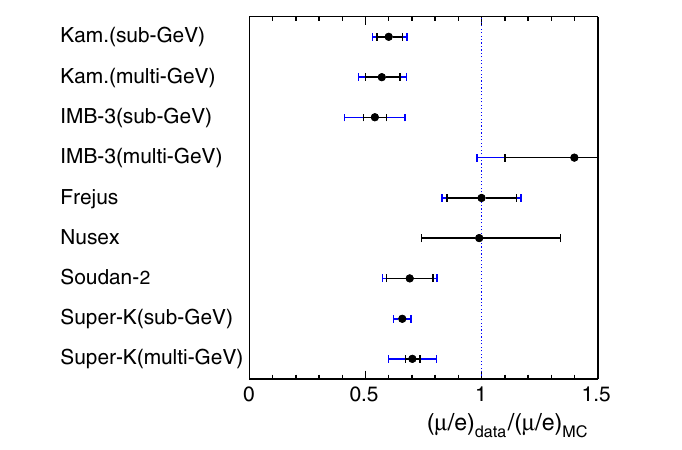}
\caption{Summary of the measurements of the double ratio $\mu$-like/e-like (data)
to $\mu$-like/e-like (Monte Carlo) by various atmospheric neutrino experiments \cite{KAJITA2004}.
If the ratio deviates from 1, it means a deficit of atmospheric muon neutrinos.
The inner and outer error bars show the statistical and statistical + systematic errors,
respectively. Around the middle of 1990s, only top six results in the figure had been published.
Obviously, only results from water Cherenkov detectors, Kamiokande and IMB,
show an atmospheric muon neutrino deficit.
}
\label{doubleratio}
\end{center}
\end{figure}

Motivated by the Kamiokande observation, other underground experiments
examined the atmospheric muon neutrino deficit.
The IMB experiment\cite{IMB1,IMB2}, which is another water Cherenkov detector,
reported results consistent with Kamiokande. However, two
tracking type experiments, Nusex \cite{NUSEX} and Frejus \cite{FREJUS}
could not find any anomaly, and their results were inconsistent with
Kamiokande. A summary of the results from various experiments is shown in Fig. \ref{doubleratio},
which was given in a review article \cite{KAJITA2004} written in 2004.

In the middle of 1990s, the Kamiokande results were not widely accepted
because of the following reasons.

\begin{enumerate}

\item The e/$\mu$ identification capability of the water
Cherenkov detectors was suspicious because only two water Cherenkov
detectors claimed an atmospheric muon neutrino deficit.
Tracking type detectors could not find the anomaly.

\item The number of total atmospheric neutrino events was $\sim$1000.
The statistics of the data were obviously poor. Much more data were definitely needed.

\item Atmospheric neutrino flux had large uncertainty. Their systematic errors
were not sufficiently small for studies of neutrino oscillations.

\item Independent confirmation by completely different type of experiment
was required.

\end{enumerate}

From middle of the 1990s, experimental efforts to settle these problems one by one
were started. Especially, long-baseline neutrino-oscillation experiments, the
main subject of this article, contributed for the third and fourth problems.

\section{E261A experiment (1992-1994)}

The e/$\mu$ identification in water Cherenkov detectors played a crucial role in the analysis
of the muon neutrino deficit.
However, the e/$\mu$ identification had been examined only by
Monte Carlo events. The E261A experiment \cite{E261A} in KEK Proton Synchrotron
(KEK-PS) was executed between 1992 and 1994. It was a “beam test” of the water
Cherenkov detector for a verification of the particle identification capability.

\begin{figure}[t!]
\begin{center}
\includegraphics[height=8cm]{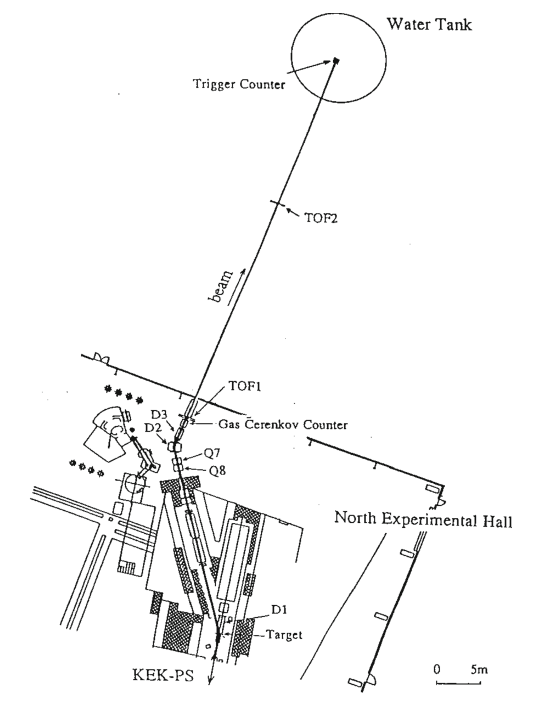}
\caption{A schematic view of the KEK-PS E261A beamline in the KEK North Counter Hall. D (dipole magnet), Q (quadrupole magnet), TOF (time of flight counter).
}
\label{E261Abeamline}
\end{center}
\end{figure}
\begin{figure}[b!]
\begin{center}
\includegraphics[height=5.5cm]{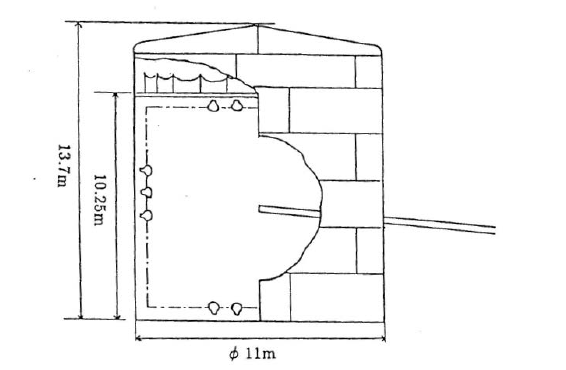}
\includegraphics[height=5cm]{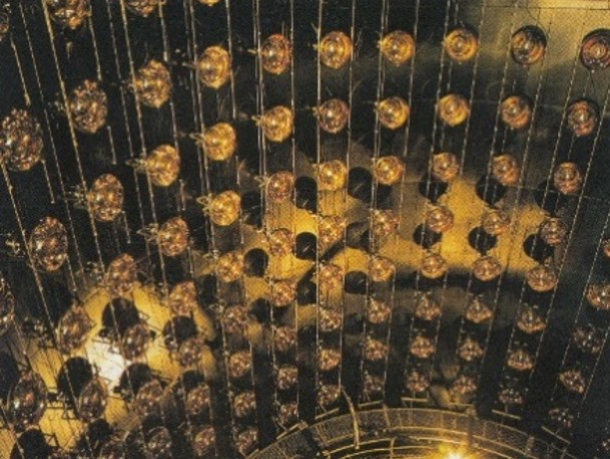}
\caption{
The vertical view of the 1-kton water Cherenkov detector (left)
and a photo of the inside of the tank(right).
}
\label{E261Atank}
\end{center}
\end{figure}

An 1-kton water Cherenkov detector, which was a miniature of
the 3-kton Kamiokande detector, was built in the KEK's North Counter Hall.
Cherenkov lights in the tank water were viewed by 380 20-inch$\Phi$ photomultiplier
tubes (PMTs) facing inward.
Schematic views of the beamline and the detector are shown in 
Figs. \ref{E261Abeamline} and \ref{E261Atank}~(left).
A photo of the inside of the tank is also shown in Fig. \ref{E261Atank}~(right).

A pipe from the North Counter Hall was inserted to the tank as shown in
Fig. \ref{E261Atank}~(left).
Electrons or muons from the KEK-PS were injected along the beam pipe.
Charged particles which travel in the pipe
were observed in the water Cherenkov detector just after they exited
from the end from the pipe. They had the same geometry as neutrino interactions
at the end of the pipe.  

The particle kinds could be known by the threshold-type gas Cherenkov
detector and time-of-flight counters along the beam pipe.
The momentum was controlled by the bending magnets in the North Counter Hall. 

\begin{figure}[t!]
\begin{center}
\includegraphics[height=7cm]{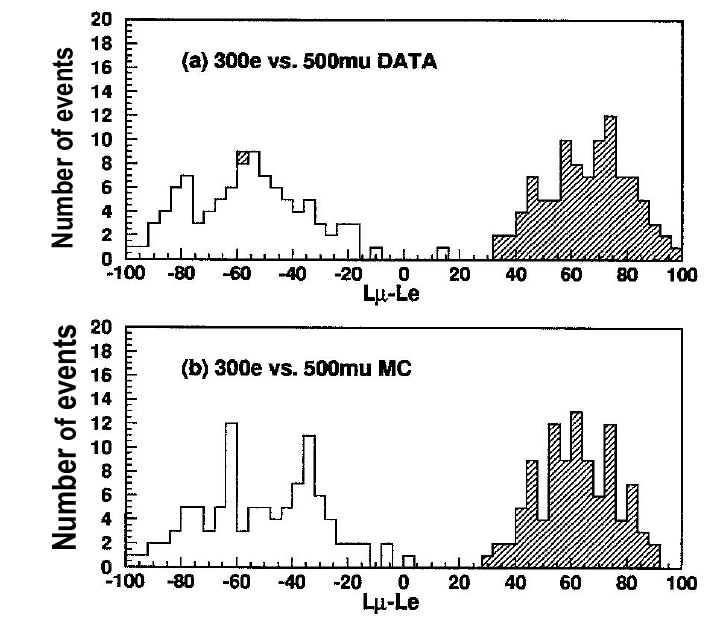}
\caption{
The $L_{\mu} - L_{e}$ distribution of (a) data and (b) Monte Carlo samples
for $e$(300 MeV/c) and $\mu$(500 MeV/c)~\cite{E261A}. 
The two data samples have about the same total photoelectrons. 
}
\label{Lm-Le}
\end{center}
\end{figure}

The same ring reconstruction program and e/$\mu$ identification algorithm as Kamiokande
were applied using charge/timing information obtained from the PMTs.
e-likelihood ($L_{e}$) and $\mu$-likelihood ($L_{\mu}$) were calculated respectively.
From a comparison between $L_{e}$ and $L_{\mu}$, particle ID was judged. 
The result of the particle identification is shown in Fig.~\ref{Lm-Le}. 
The particle
identification algorithm could clearly separate
electron beam events and muon beam events.
It was experimentally verified that the e/$\mu$ identification
capability was better than 99\%.
The possibility that the atmospheric muon neutrino deficit was
due to poor identification of the water Cherenkov
detectors was clearly excluded.

\section{Super-Kamiokande experiment (1996-~~)} 

Super-Kamiokande is a water Cherenkov detector 
located at 1000 m.w.e. underground in the Kamioka mine, Japan.
A 50-kton mass of pure water is viewed by 11146 20-inch$\Phi$ photomultipliers.
The fiducial volume is 22.5 kton. 

The first proposal of the Super-Kamiokande experiment was given
by Prof. Koshiba in December 1983.
At that time, 
the first half-year data from Kamiokande showed the nucleon lifetime is 
much longer than the theoretical expectations.
Thus, the main purpose of the experiment
was search for nucleon decay in a longer lifetime region.
After the retirement of Prof. Koshiba, Prof. Yoji Totsuka succeeded
the leader of the project. The project was approved by the Japanese
government in 1991, and the experiment started in 1996 after five years
of detector construction.

\begin{figure}[t!]
\begin{center}
\includegraphics[height=8cm]{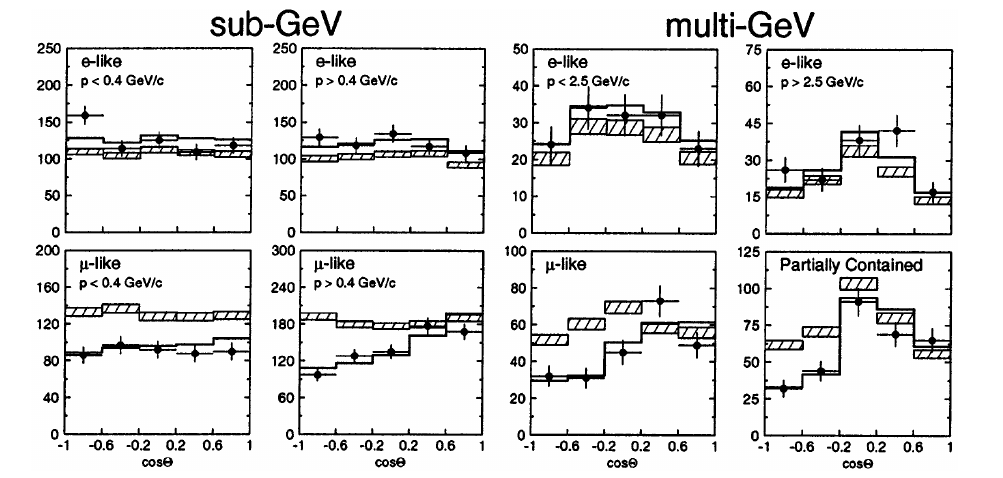}
\caption{
Zenith angle distribution of atmospheric neutrino events reported
by the Super-Kamiokande collaboration~\cite{SKATM}.
Zenith angle distributions of $\mu$-like and e-like events for sub-GeV and
multi-GeV data sets. Upward-going particles
have $\cos\Theta < 0$ and downward-going particles have $\cos\Theta  >  0$.
Sub-GeV data are shown separately for $p < 400~{\rm MeV/c}$ and $p > 400~{\rm MeV/c}$.
Multi-GeV e-like distributions are shown for $p < 2.5$ and $p > 2.5 {\rm GeV/c}$ and the
multi-GeV $\mu$-like are shown separately for fully-contained and partially-contained events.
The hatched region shows the Monte Carlo expectation for no oscillations normalized to the data
live time with statistical errors. The bold line is the best-fit expectation for
\nmnt oscillations with the overall flux normalization fitted as a free parameter.
}
\label{SKzenith}
\end{center}
\end{figure}

At the 18th International Conference on Neutrino Physics and Astrophysics (NEUTRINO 1998)
in Takayama \cite{NEUTRINO1998}, the Super-Kamiokande collaboration announced the discovery of
atmospheric neutrino oscillations with a statistical significance of 6.2$\sigma$.
Just after the conference, the results were officially published~\cite{SKATM}
based on 4654 atmospheric neutrino events accumulated in 535
days, corresponding to 33.0 kton$\cdot$yr.

In the paper, the number of electron neutrinos well agreed with expectation,
but the number of muon neutrinos was clearly smaller than the expectation
in all energy regions as shown in Fig.~\ref{SKzenith}.
Significant zenith angle distributions were also found, and
the data were consistent with two-flavor \nmnt oscillations.

\begin{figure}[t!]
\begin{center}
\includegraphics[height=9cm]{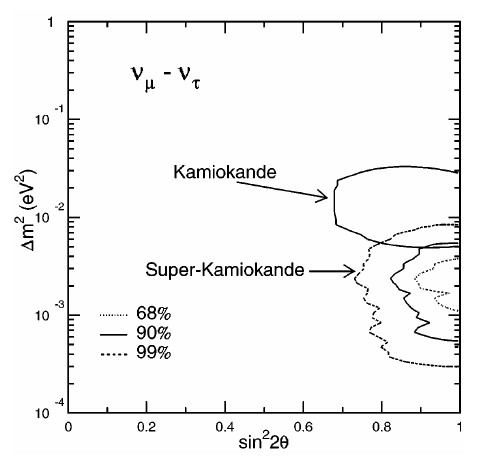}
\caption{
Constraints on neutrino oscillation parameters given in Ref.~\cite{SKATM}.
The 68\%, 90\%, and 99\% confidence intervals are
shown for $\sin^{2}2\theta$ and $\Delta m^{2}$ for \nmnt
two-neutrino oscillations based on 33.0 kt$\cdot$yr of
Super-Kamiokande data. The 90\% confidence interval obtained
by the Kamiokande experiment is also shown.
}
\label{SKcontour}
\end{center}
\end{figure}

The constraints on the neutrino oscillation parameters
($\Delta m^{2}$, $\sin^{2}2\theta$) were obtained as shown in Fig.~\ref{SKcontour}.
The data were consistent with two-flavor \nmnt oscillation with
$$\sin^{2}2\theta > 0.82  {\rm ~~~and~~~}
5\times 10^{-4} {\rm eV^{2}} < \Delta m^{2} < 6\times 10^{-3} {\rm eV^{2}}$$
at 90\% confidence level.
The best fit parameters are
$$\sin^{2}2\theta = 1.0 {\rm ~~~and~~~} \Delta m^{2} = 2.2\times 10^{-3} {\rm eV^{2}}$$
After more than 20 years, this parameter region still agrees well with the updated numbers;
$$\sin^{2}\theta = 0.545 \sim 0.547 {\rm ~~~and~~~}
|\Delta m^{2}| = (2.453\sim 2.546)\times 10^{-3} {\rm eV^{2}},$$
which are given in PDG 2020~\cite{PDG2020}. 

\section{K2K experiment (1999-2004)} 

After the drastic improvement of the statistics of the neutrino events
in Super-Kamiokande,
the remaining key issue was the neutrino source used for experiments.
The calculation of atmospheric neutrino flux was based on
many ambiguous measurements. They are primary proton (and heavier nuclei) fluxes,  
geomagnetic field including effects from solar activity,
hadronic interaction models, and atmospheric temperature.
All of these measurements had large uncertainties~\cite{HONDA1}. 
The atmospheric muon neutrino deficit papers claimed that even though
absolute atmospheric neutrino fluxes
have large uncertainty, the $\nue$/$\num$ ratio is robust.
However, it is obvious that atmospheric neutrinos were not well
understood for the final confirmation of neutrino oscillations.
In addition, high energy physicists strongly hope to
examine neutrino oscillations using their own neutrino beam.
Artificial neutrino beam can be controlled by themselves, and
the neutrino fluxes just after their production
can be measured by physicists directly.

\begin{figure}[t!]
\begin{center}
\includegraphics[height=9cm]{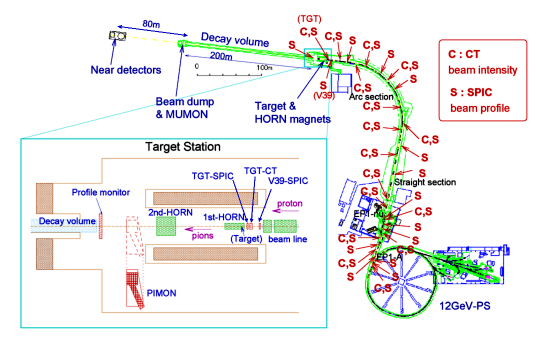}
\caption{
A schematic view of the KEK-PS and
neutrino beam line and the location of beam line components.
The neutrino beam line leads protons through a distance of
400 m from the extraction point to the target station via
the straight and arc sections. The characters ‘‘C’’ and ‘‘S’’ in the
figure show the locations of the CT and SPIC installations,
respectively. The lower-left inset is a magnified view of the
target station. The production target and a set of horn magnets
are located in the target station. A pion monitor was installed on
two occasions downstream of the horn magnets. MUMON (muon monitor),
PIMON (pion monitor), TGT (target).
}
\label{K2Kbeamline}
\end{center}
\end{figure}

The idea of a long-baseline neutrino-oscillation experiment
with an accelerator neutrino beam was presented by
Prof. M.~Koshiba at “A workshop for High Intensity facility
in 1988” at Breckenridge, Colorado~\cite{LBKOSHIBA}.
This workshop was
primarily concerned with the physics program at Fermilab in
1990, using the high intensity Main Injector accelerator. He
discussed a large water Cherenkov detector as a far detector
of a long-baseline experiment with the FNAL Main Injector.
 
The first long-baseline neutrino-oscillation experiment
was K2K (KEK to Kamioka) experiment started in 1999~\cite{K2K}.
A muon neutrino beam generated at KEK was shot toward the Super-Kamiokande
detector located 250~km away from KEK.
The purpose of the experiment was an examination of the \nmnt oscillation
reported by Kamiokande and Super-Kamiokande using
an artificial neutrino beam.

The configuration of KEK-PS and
the neutrino beamline is given
in Fig.~\ref{K2Kbeamline}.
The proton beam produced by KEK-PS was
bent in the arc section toward the direction of Super-Kamiokande.
The proton beam was 1.1-$\mu$sec beam duration
in a 2.2-s accelerator cycle. The proton intensity was
 $(5\sim 6) \times 10^{12}$ proton per pulse (ppp).

The protons hit an aluminum target in the target station.
Positively charged particles, mainly pions, were produced,
and they were focused to the direction of the Super-Kamiokande
detector by a pair of pulsed magnetic horns.
The neutrinos produced from the decays of these particles
in the 200~m of decay volume
were $\sim$99\% pure muon neutrinos with a mean energy of 1.3 GeV.

\begin{figure}[t!]
\begin{center}
\includegraphics[height=8cm]{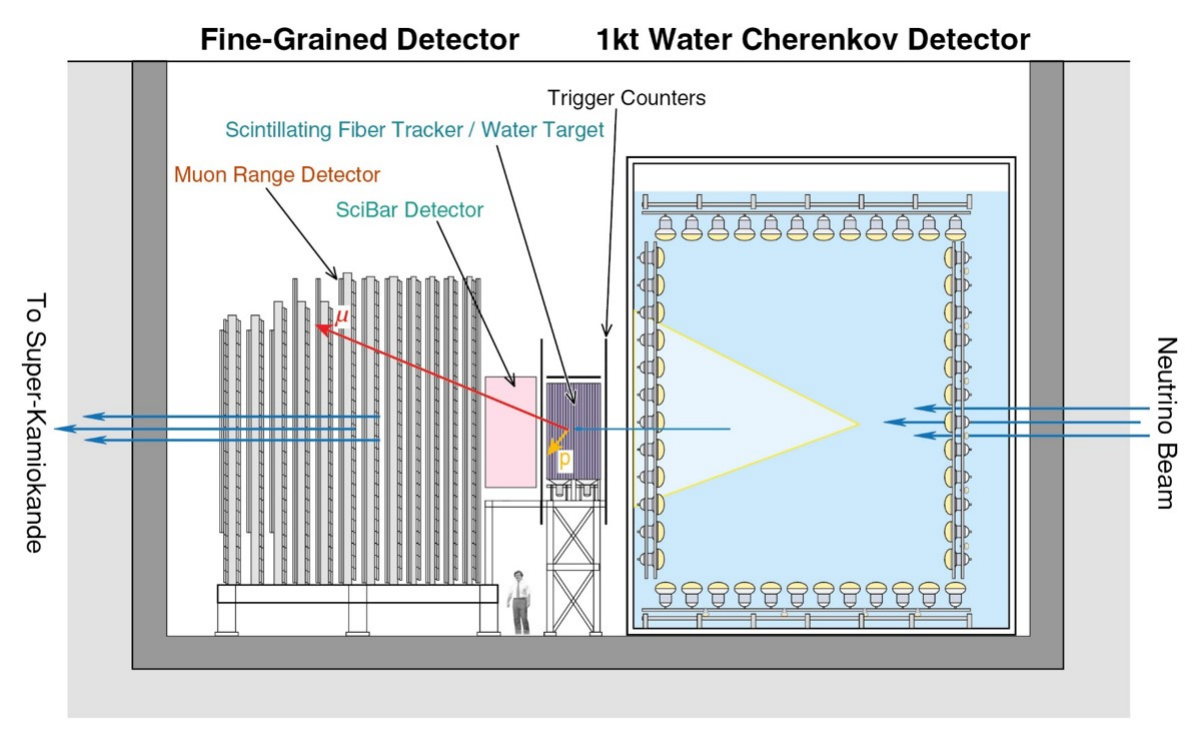}
\caption{
K2K near detector
}
\label{K2Knear}
\end{center}
\end{figure}

Two near detectors were constructed 300~m downstream
of the target. They were the “one kton water Cherenkov detector (1kt)” and
“Fine-Grained detector (FGD)”. The schematic view of the near
detectors is shown in Fig.~\ref{K2Knear}.

The 1kt detector was a 1/50 miniature of the Super-Kamiokande detector.
Since the experimental technology was exactly the same as that of the far detector,
systematic errors related to the detection method were cancelled out,
and direct comparisons of the data between the near detector and the far detector
were possible.
The FGD consisted of four detector components. They were a scintillating fiber tracker
with water target, trigger counter, lead glass counters, and muon range detector.
The lead glass counters were replaced with a SciBar detector in the second
half of the experiment.
The purpose of the FGD was to obtain precise measurements of neutrino beam properties.


From the analysis of the near detector data, many properties of the
neutrino beam were experimentally obtained.
Firstly, the energy spectrum of the beam was precisely calculated.
The mean energy of the neutrino beam was about 1.3~GeV, and the
peak energy was about 1.0 GeV. The beam was almost a pure muon neutrino
beam, and the $\nue/\num$ is $\sim$1\%.
The beam direction was adjusted within 1~mrad.
The neutrino flux and energy spectrum are nearly identical within 3~mrad.
Since the angular size of the Super-Kamiokande detector from
KEK was $\sim$50m/250km = 0.2~mrad, the systematic errors on the
neutrino beam properties due to the beam direction were negligible.
From the extrapolation of the neutrino flux in the near detectors,
the neutrino flux at Super-Kamiokande was expected to be
$1.3\times 10^{6} \nu/{\rm cm^{2}}$ for $10^{20}$ POT~(Proton On Target).
It corresponds to $\sim$170 neutrino events in the 22.5~kton of
fiducial volume of Super-Kamiokande in the case of null oscillation.


The experiment started in June 1999 and ended in November 2004.
Between November 2001 and the end of 2002, the experiment stopped
because of trouble of the Super-Kamiokande detector.
The record of the beam in the experiment is summarized
in Table~\ref{tab:K2Ksummary} 

\begin{table}[t!]
\caption{Summary of the beam operation in the K2K experiment}
\begin{center}
\begin{tabular}{ll}
\hline
\hline
~~ Beam period             & Jun~~4, 1999 - Jul~12, 2001 (K2K-I)\\        
                        & Jan~17, 2003 - Nov~~6, 2004 (K2K-II)\\
~~ Total beam time         & 442.8~days (233.7 for K2K-I + 209.1 for K2K-II)\\
~~ Total spill numbers     & $17.4 \times10^{6}$ spills\\
~~ Total POT for analysis & $92.2 \times 10^{18}$\\
\hline
\hline
\end{tabular}
\end{center}
\label{tab:K2Ksummary}
\end{table}

The criteria of the KEK beam neutrino event selection in the 
Super-Kamiokande detector were almost the same as those for atmospheric
neutrino events. One additional condition was the coincidence
with the neutrino beam period because the neutrino beam period
was only 1.1-$\mu$s in a 2.2-s accelerator cycle.
The absolute beam time at the neutrino beamline in KEK and event time at
Super-Kamiokande were individually recorded using GPS systems.
The events in the 1.5-$\mu$s time window in accordance with the
beam period were selected.

\begin{figure}[t!]
\begin{center}
\includegraphics[height=8cm]{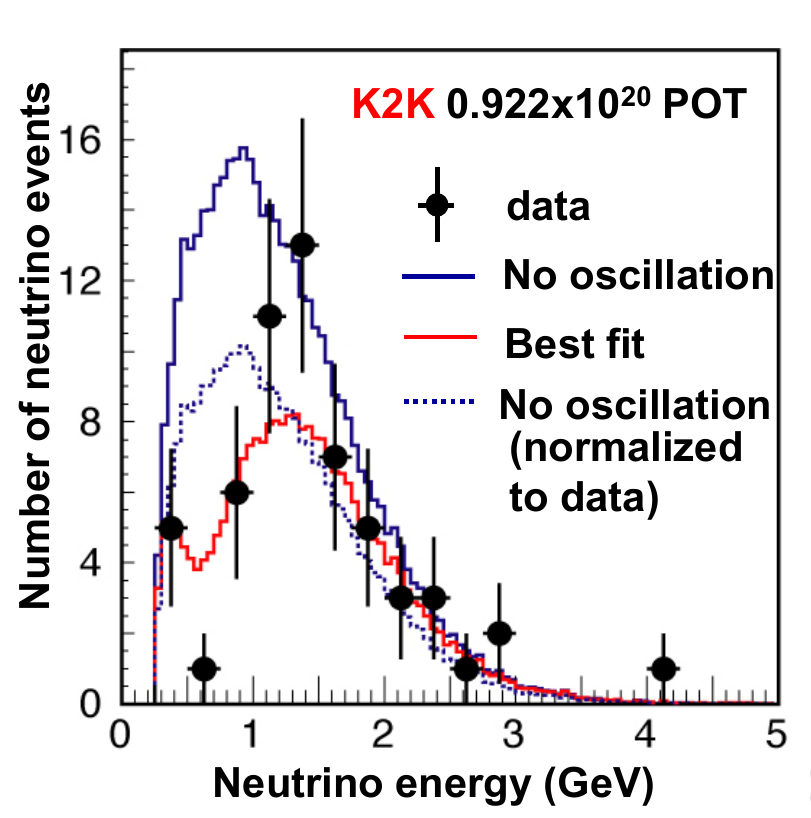}
\caption{
Neutrino energy spectrum obtained from 112 single-ring $\mu$-like events
detected from $0.922\times 10^{20}$ proton on target in the K2K experiment.
The data are shown by filled circles with error bar, and the expectation for
null oscillation is shown by solid blue line. The distortion of the energy spectrum
in the 0.5-1.0 GeV range is evidence of neutrino oscillation.
}
\label{K2Kspectrum}
\end{center}
\end{figure}


From a total of $0.922\times 10^{20}$ POT data, 112 neutrino events were obtained.
The expected number of neutrino events was calculated from extrapolation of the events
in the near detectors, and was obtained to be $158.1^{+9.2}_{-8.6}$. The observation was
considerably smaller than the expectation.


Among the 112 neutrino events, 58 events were single ring $\mu$-like events, which were
thought to be generated by quasi-elastic charged current interaction,
$$\nu_{\mu} + n \rightarrow \mu^{-} + p$$
For these events, neutrino energy ($E_{\nu}$) can be calculated from the muon energy
and opening angle from the neutrino travel direction ($\theta_{\mu}$).
$$E_{\nu}={{m_{N} E_{\mu}-m^{2}_{\mu}/2}\over{m_{N}-E_{\mu}+p_{\mu} \cos\theta_{\mu}}}.$$
The neutrino energy spectrum is shown in Fig.~\ref{K2Kspectrum}.
The result showed a clear distortion in the 0.5-1.0 GeV range,
which reflects the energy dependence of neutrino survival probability.
Both the total number of neutrino events and the shape of the neutrino
energy spectrum showed clear disagreement with the expectation for null oscillation.
If null oscillation was assumed, such poor agreements happened
with a probability of 0.0015\%(4.3$\sigma$).

\begin{figure}[t!]
\begin{center}
\includegraphics[height=8cm]{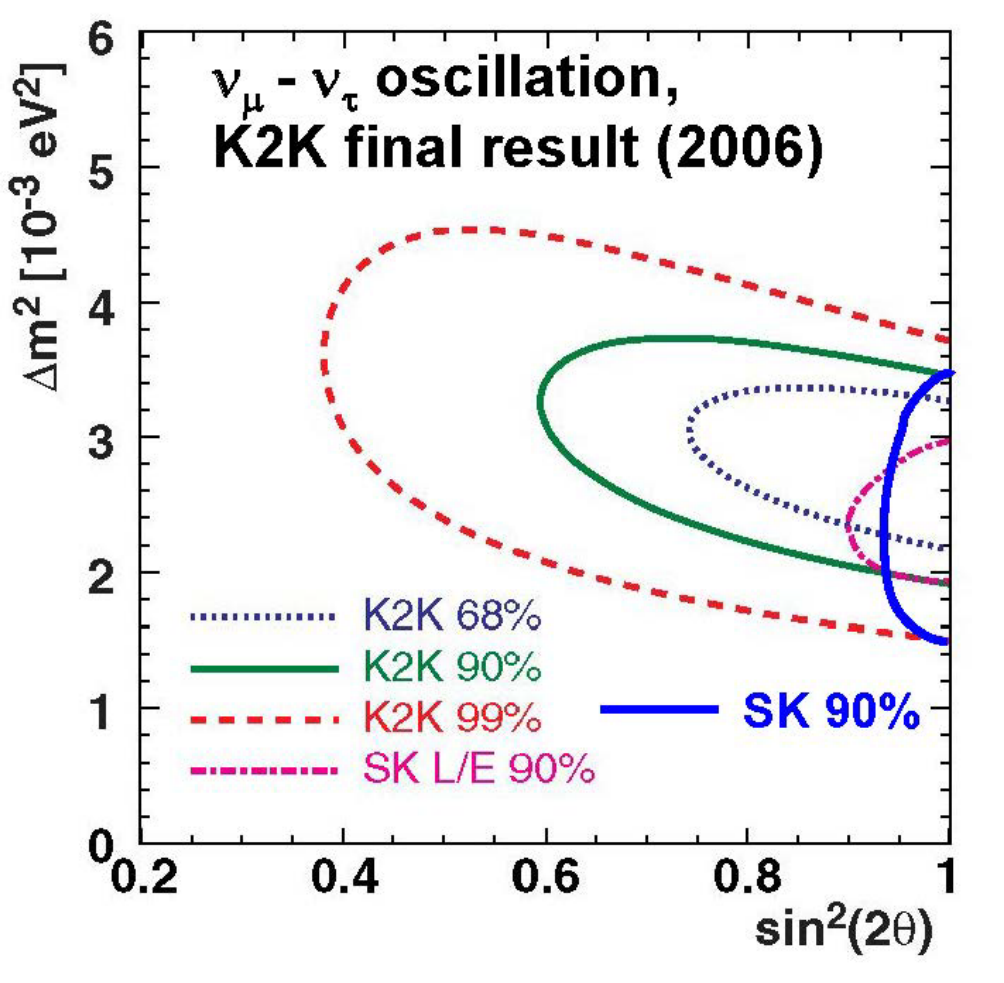}
\caption{
Constraints for \nmnt neutrino oscillation parameter region
in the $(\Delta {\rm m^{2}}, \sin^{2}2\theta)$ plane
by the K2K experiment. The constraints from the Super-Kamiokande
experiment are also shown.
}
\label{K2Kcontour}
\end{center}
\end{figure}


Constraints for \nmnt neutrino oscillation parameter region were calculated
in the $(\Delta {\rm m^{2}}, \sin^{2}2\theta)$ plane. The result is shown in
Fig.~\ref{K2Kcontour}.
The best fit parameters are
$$\sin^{2}2\theta = 1.0 {\rm ~~~and~~~}
\Delta m^{2} = 2.8\times 10^{-3} {\rm eV^{2}}.$$
For the maximal oscillation, $\sin^{2}2\theta=1$, the 90\% confidence level
interval for $\Delta m^{2}$ is
$$\Delta m^{2} = (1.9\sim 3.5)\times 10^{-3} {\rm eV^{2}}.$$
The constraints agreed with atmospheric neutrino results
from Super-Kamiokande.
This is the first confirmation of \nmnt oscillation by an
artificial neutrino beam.

\section{MINOS experiment (2005-2012)}

Although the K2K experiment confirmed \nmnt neutrino oscillation
by an artificial neutrino beam, it was not recognized as an experiment that was
perfectly independent from Super-Kamiokande. This was because the K2K experiment
used Super-Kamiokande as the far detector, and the members of collaborations had a large
overlap with each other.

The first completely independent experiment was the MINOS experiment~\cite{MINOS}.
The MINOS experiment was a long-baseline neutrino-oscillation experiment using the FNAL
120 GeV Main Injector (NuMI) and far detector in Soudan Mine, 735 km away.
The far detector was a “sandwich” of iron plate of 2.54 cm thickness and scintillator of 1.00 cm thickness.
The total volume was 5.4 kton. They were in 1.3 Tesla of magnetic field.
A 0.98-kton near detector of similar design was also constructed in FNAL.

\begin{figure}[t!]
\begin{center}
\includegraphics[height=8cm]{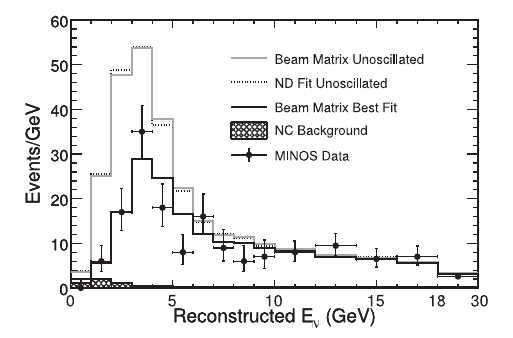}
\caption{
Comparison of the FD (Far Detector) $E^{\rm reco}_{\nu}$ spectrum with predictions
for no oscillations for both analysis methods and for oscillations
with the best-fit parameters from the beam matrix extrapolation
from ND (Near Detector)~\cite{MINOS}. The estimated NC (Neutral Current) background is also shown.
The last energy bin contains events between 18 and 30 GeV.
}
\label{MINOSspectrum}
\end{center}
\end{figure}

The MINOS experiment started in 2005 and published the first result of muon neutrino
disappearance in 2006.
From $1.27\times10^{20}$ POT data, 215 muon neutrino events were found, where 336$\pm$14
events were expected if no oscillation was assumed.
A distortion of the energy spectrum which was a strong
indication of neutrino oscillation was also found, as shown in Fig.~\ref{MINOSspectrum}.

The data were consistent with muon neutrino disappearance via oscillations with

$$\sin^{2}2\theta_{23}~>~0.87~~~(68\%~{\rm C.L.}) {\rm ~~~and~~~}
\Delta m^{2}_{32}	= (2.74^{+0.44}_{-0.27}) \times 10^{-3} {\rm ~eV^{2}}$$

\noindent
The result was consistent with Super-Kamiokande and K2K,
and was the first perfectly independent and convincing confirmation
of the \nmnt neutrino oscillation discovered by Super-Kamiokande.
The independence includes different detector type, different neutrino source,
and different collaboration members.
 
\section{Summary and beyond}

The atmospheric muon neutrinos deficit observed by Kamiokande was thought to be
evidence of \nmnt neutrino oscillation. The evidence was experimentally examined and
\nmnt oscillation was
finally confirmed by contributions from many experiments, E261A, Super-Kamiokande,
K2K, and MINOS.
It was widely recognized that \nmnt oscillation was discovered by Super-Kamiokande
experiment because of its large event numbers and sufficient statistical significance.
The Nobel Prize in Physics 2015 was awarded to Prof. T. Kajita of the Super-Kamiokande
collaboration “for the discovery of neutrino oscillations, which shows that neutrinos have mass”.

In the same time duration as the discovery of \nmnt oscillation,
the neutrino oscillation between $\nue$ and $\num$
was also discovered from observations of solar neutrinos~\cite{NAKAHATA}.
There were large contributions from Homestake~\cite{HOMESTAKE},
Kamiokande~\cite{KAMSOLAR}, Super-Kamiokande~\cite{SKSOLAR}, and SNO~\cite{SNO}.

After independent discoveries of \nmnt and \nenm oscillations,
study of neutrino oscillations was extended to three flavor oscillations~\cite{THREEFLAVOR}.
In the framework of three flavor oscillations, the mass matrix, which expresses
the correlation between the flavor eigenstates and the mass eigenstates, has
six parameters. Among these six parameters, 
the remaining unknown parameters were mixing
angle between first and third generation, $\theta_{13}$, and the CP
violation phase, $\delta_{\rm CP}$.

Determination of the remaining unknown parameters became the most important
research subject of neutrino experiments from the
middle of the 2000s.
The appearance of electron neutrinos in the muon neutrino beam, which is
evidence of finite $\theta_{13}$, was found by the T2K experiment~\cite{T2KTHETA13}.
Possible evidence of the CP violation phase was found from the difference of the
oscillation probabilities between neutrino and anti-neutrino also by
the T2K experiment~\cite{T2KDCP}. There were also contributions from
NOvA experiment~\cite{NOVA}, which is another long-baseline neutrino-oscillation
experiment in the United States. 
The Hyper-Kamiokande experiment~\cite{HYPERK} and DUNE
experiment~\cite{DUNE} for the precise measurement
of the CP violation phase are in preparation.
However, these advances are outside of the scope of this document,
and are not described in this article.

The evidences for neutrino oscillations, both \nmnt and \nenm, observed
by Kamiokande led to many discoveries and progresses in neutrino physics.
Pioneering contributions from Prof. M. Koshiba should be emphasized.

\section*{Acknowledgments}
The author would like to acknowledge the colleagues in the Kamiokande, E261A, Super-Kamiokande,
K2K, and T2K collaborations.
The author also thanks Profs M. Nozaki and M. Yokoyama for giving him
the opportunity to write this article.
\medskip
\let\doi\relax


\begin{thebibliography}{99}
\bibitem{KAMATM1}
  K. S. Hirata et al. (Kamiokande collaboration), Phys. Lett. {\bf B205}, 416(1988).
\bibitem{KAMATM2}
  K. S. Hirata et al. (Kamiokande collaboration), Phys. Lett. {\bf B280}, 146(1992).
\bibitem{KAMATM3}
  K. S. Hirata et al. (Kamiokande collaboration), Phys. Lett. {\bf B335}, 237(1994).
\bibitem{KAJITA2023}
  T. Kajita, Prog. Theor. Exp. Phys. {\bf 2023}, \\
  DOI: 10.1093/ptep/ptad013
\bibitem{SUZUKI}
  A. Suzuki, Prog. Theor. Exp. Phys. {\bf 2022}, 12B102(2022).
\bibitem{IMB1}
  D. Casper et al. (IMB collaboration), Phys. Rev. Lett. {\bf 66}, 2561(1991).
\bibitem{IMB2}
  R. Becker-Szendy et al. (IMB collaboration), Phys. Rev. {\bf D46}, 3720(1992).
\bibitem{NUSEX}
  M. Aglietta et al. (Nusex collaboration), Europhys. Lett. {\bf 8}, 611(1989).
\bibitem{FREJUS}
  Ch. Berger et al. (Frejus collaboration), Phys. Lett. {\bf B227}, 489(1989).
\bibitem{KAJITA2004}
  T. Kajita, New Journal of Physics {\bf 6}, 194(2004).      
\bibitem{E261A}
  S. Kasuga et al. (E261A collaboration), Phys. Lett. {\bf B374}, 238(1996).
\bibitem{NEUTRINO1998}
  T. Kajita for Super-Kamiokande collaboration,\\
  Proceedings of International Conference on Neutrino physics and astrophysics (Neutrino 1998),\\ 
  Nucl. Phys. B Proc.Suppl. {\bf 77}, 123(1999).
\bibitem{SKATM}
  Y. Fukuda et al. (Super-Kamiokande collaboration), Phys. Rev. Lett. {\bf 81}, 1562(1998).
\bibitem{PDG2020}
  P. A. Zyla et al. (Particle Data Group), Prog. Theor. Exp. Phys. {\bf 2020}, 083C01(2020).
\bibitem{HONDA1}
  M. Honda et al., Phys. Lett. {\bf B248}, 193(1990).\\
  M. Honda et al., Phys. Rev. {\bf D52}, 4985(1995).
\bibitem{LBKOSHIBA}
  M. Koshiba, Summary report on long baseline neutrino oscillation,
  in "Workshop on Physics at the Main Injector (to be followed  
  by Fermilab Annual Users Meeting May 19–20)", 139-145,
  Batavia, Illinois, May 16-18, 1989.~~
  https://inspirehep.net/conferences/967090
\bibitem{K2K}
  E. Aliu et al. (K2K collaboration), Phys. Rev. Lett. {\bf 94}, 081802(2005).\\
  M. H. Ahn et al. (K2K collaboration), Phys. Rev. {\bf D74}, 072003(2006).
\bibitem{MINOS}
  D. G. Michael et al. (MINOS collaboration), Phys. Rev. Lett. {\bf 97}, 191801(2006).
\bibitem{NAKAHATA}
  M. Nakahata, Prog. Theor. Exp. Phys. {\bf 2022}, 12B103(2022).
\bibitem{HOMESTAKE}
  B. T. Cleveland et al. (Homestake collaboration), Astrophys. J. {\bf 496}, 505(1998).
\bibitem{KAMSOLAR}
  K. S. Hirata et al. (Kamiokande collaboration), Phys. Rev. Lett. {\bf 63}, 16(1989).\\
  Y. Fukuda et al. (Kamiokande collaboration), Phys. Rev. Lett. {\bf 77}, 1683(1996).
\bibitem{SKSOLAR}
  Y. Fukuda et al. (Super-Kamiokande collaboration), Phys. Rev. Lett. {\bf 81}, 1158(1998).
\bibitem{SNO}
  Q. R. Ahmad et al. (SNO collaboration), Phys. Rev. Lett. {\bf 89}, 011301(2002).
\bibitem{THREEFLAVOR}
  Z. Maki, M. Nakagawa, and S. Sakata, Prog. Theor. Phys. {\bf 28}, 870(1962).\\
  B. Pontecorvo, Sov. Phys. JETP {\bf 26}, 984(1968).
\bibitem{T2KTHETA13}
  K. Abe et al. (T2K collaboration), Phys. Rev. Lett. {\bf 107}, 041801(2011).\\
  K. Abe et al. (T2K collaboration), Phys. Rev. Lett. {\bf 112}, 061802(2014).
\bibitem{T2KDCP}
  K. Abe et al. (T2K collaboration), Phys. Rev. Lett. {\bf 118}, 151801(2017).\\
  K. Abe et al. (T2K collaboration), Phys. Rev. Lett. {\bf 121}, 171802(2018).\\
  K. Abe et al. (T2K collaboration), Nature {\bf 580}, 339(2020).
\bibitem{NOVA}
  P. Adamson et al. (NOvA collaboration), Phys. Rev. Lett. {\bf 116}, 151806(2016).\\
  M. A. Acero et al. (NOvA collaboration), Phys. Rev. Lett. {\bf 123}, 151803(2019).
\bibitem{HYPERK}
  K. Abe et al. (Hyper-Kamiokande collaboration), Prog. Theor. Exp. Phys. {\bf 2015}, 053C02(2015).
\bibitem{DUNE}
  B. Abi et al. (DUNE Collaboration), J. Instrum. {\bf15}, T08008 (2020).
\end{thebibliography}
\end{document}